\documentclass[aps, prl, twocolumn, groupedaddress, showpacs, showkeys]{revtex4}
\usepackage{graphicx}
\usepackage{epstopdf}
\usepackage{dcolumn}
\usepackage{amsmath,amssymb}

\def\bx{{\mathbf{x}}}

\newcommand{\beq}{\begin{equation}}
\newcommand{\eeq}{\end{equation}}
\newcommand{\bea}{\begin{eqnarray}}
\newcommand{\eea}{\end{eqnarray}}
\newcommand{\ba}{\begin{array}}
\newcommand{\ea}{\end{array}}
\newcommand{\bit}{\begin{itemize}}
\newcommand{\eit}{\end{itemize}}
\newcommand{\eq}{Eq.~}

\newcommand{\fig}{Fig.~}

\def\math{\mathsurround 0pt}
\def\oversim#1#2{\lower.5pt\vbox{\baselineskip0pt \lineskip-.5pt
        \ialign{$\math#1\hfil##\hfil$\crcr#2\crcr{\scriptstyle\sim}\crcr}}}

\def\lsi{\raise0.3ex\hbox{$<$\kern-0.75em\raise-1.1ex\hbox{$\sim$}}}
\def\gsi{\raise0.3ex\hbox{$>$\kern-0.75em\raise-1.1ex\hbox{$\sim$}}}
\newcommand{\lsim}{\mathop{\lsi}}

\begin{document}

\title{\vspace*{-1.5cm}
\begin{flushright}
\texttt{\footnotesize CERN-PH-TH/2010-079}
\end{flushright}
\vfill

Constraining the QCD phase diagram by 
tricritical lines \\at imaginary chemical potential}
\author{Philippe de Forcrand$^{1,2}$}
\email{forcrand@phys.ethz.ch}
\author{Owe Philipsen$^{3}$}
\email{philipsen@th.physik.uni-frankfurt.de}
\affiliation{$^1$Institut f\"ur Theoretische Physik, ETH Z\"urich, CH-8093 Z\"urich, Switzerland\\
$^2$CERN, Physics Department, TH Unit, CH-1211 Geneva 23,
Switzerland\\
$^3$Institut f\"ur Theoretische Physik, Johann-Wolfgang-Goethe-Universit\"at,
                   60438 Frankfurt am Main, Germany }

\date{\today}

\begin{abstract}
We present unambiguous evidence from lattice simulations of QCD with three degenerate
quark species for two tricritical points in the $(T,m)$ phase diagram at fixed imaginary $\mu/T=i\pi/3$ 
mod $2\pi/3$,
one in the light and one in the heavy mass regime. These represent the boundaries of the chiral and deconfinement critical lines continued to imaginary chemical potential, respectively. 
It is demonstrated that the shape of the deconfinement critical line for real chemical potentials is dictated
by tricritical scaling and implies the weakening of the deconfinement transition with real chemical 
potential. The generalization to non-degenerate and light quark masses is discussed. 

\end{abstract}


\maketitle


The QCD phase diagram is at present largely unknown.
It describes which different forms of nuclear matter exist for 
different choices of temperature and baryon 
density, and whether they are separated by phase transitions.
Its knowldedge is thus of great importance for current and future 
experimental programs in nuclear and heavy ion physics
as well as astro-particle physics. Since QCD is strongly coupled on scales of a baryon
mass and below, fully non-perturbative calculations
are warranted.   
Unfortunately, Monte Carlo simulations of lattice QCD 
at non-vanishing baryon density
are prohibited by the ``sign'' problem.
To date only indirect methods are available,
introducing additional approximations 
which are justified for $\mu/T\lsim 1$ only \cite{review}. 
One of these consists
of simulating QCD at imaginary chemical potential $\mu=i\mu_i$ with $\mu_i\in \mathbb{R}$, 
for which there is no sign problem, 
and analytically continuing the results to real chemical potential \cite{fp1,el1}.
While the Monte Carlo results contain the full information about imaginary $\mu$,
analytic continuation via truncated polynomials fitted to the data 
introduces the approximation.

In this letter we propose instead to study 
the phase diagram of QCD at imaginary chemical potential in its own right. 
We shall demonstrate that
there are intricate first order, triple, critical and tricritical structures,
whose details depend on the number of dynamical quark flavours $N_f$ 
and their respective masses $m_f$. These structures are {\it bona fide} properties 
of QCD and for this reason alone
merit a detailed investigation. Moreover, we show that
tricritical lines found at imaginary chemical potential, with their
associated scaling behaviour, represent important constraints for the critical 
surfaces at real chemical potential. 
Finally, the phase diagram we investigate may serve as 
benchmark for studies within effective models (such as PNJL, 
sigma models, quark hadron models etc.), which 
can be easily extended to imaginary $\mu$.

Here we present a study of the $(T,m)$ phase structure of QCD
at fixed imaginary chemical potential $(\mu/T)_c=i(2n+1)\pi T/3, n=0,\pm 1,\pm 2,\ldots$
for $N_f=3$ degenerate quark flavours. 
At those values of $\mu$, QCD undergoes a  
transition between adjacent $Z(3)$ sectors. This is due to the exact symmetries of the partition
function, 
\beq
Z(\mu)=Z(-\mu), \quad 
Z\left(\frac{\mu}{T}\right) = Z\left(\frac{\mu}{T}+i\frac{2\pi n}{3}\right)\;,
\label{zsym}
\eeq
for complex $\mu$ \cite{rw}.
The different $Z(3)$-sectors can be distinguished by 
the Polyakov loop 
\beq
L(\bx)=\frac{1}{3}{\rm Tr} \prod_{\tau=1}^{N_\tau}U_0(\bx,\tau)=|L|\,{\rm e}^{-i\varphi}\;,
\eeq
whose phase $\varphi$ cycles through $\langle \varphi \rangle =n (2\pi/3), n=0,1,2,\ldots$
as the different sectors are traversed.

\begin{figure}[t]
\hspace*{-0.5cm}
\includegraphics[width=0.24\textwidth]{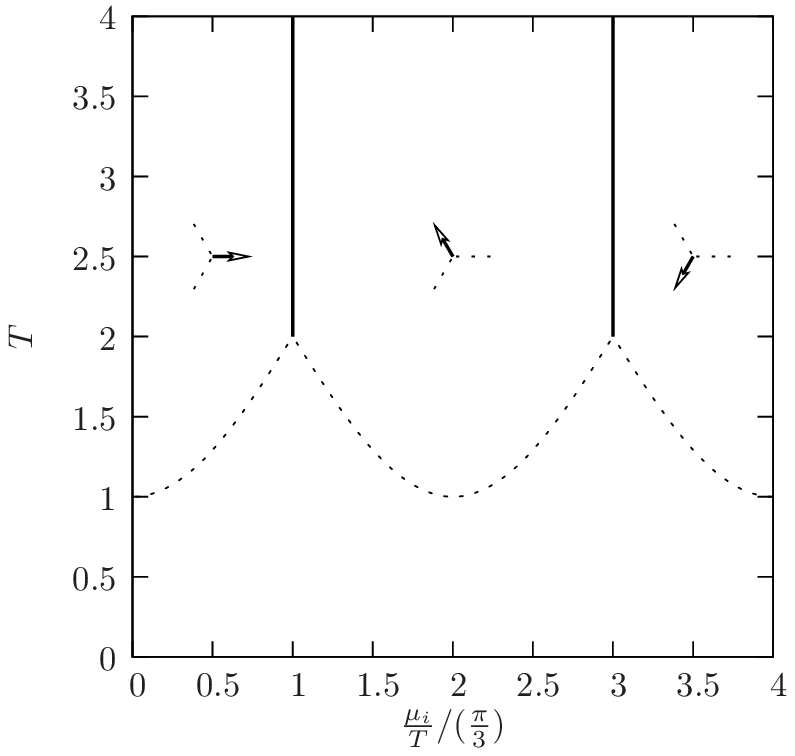}\hspace*{0.2cm}
\includegraphics[width=0.22\textwidth]{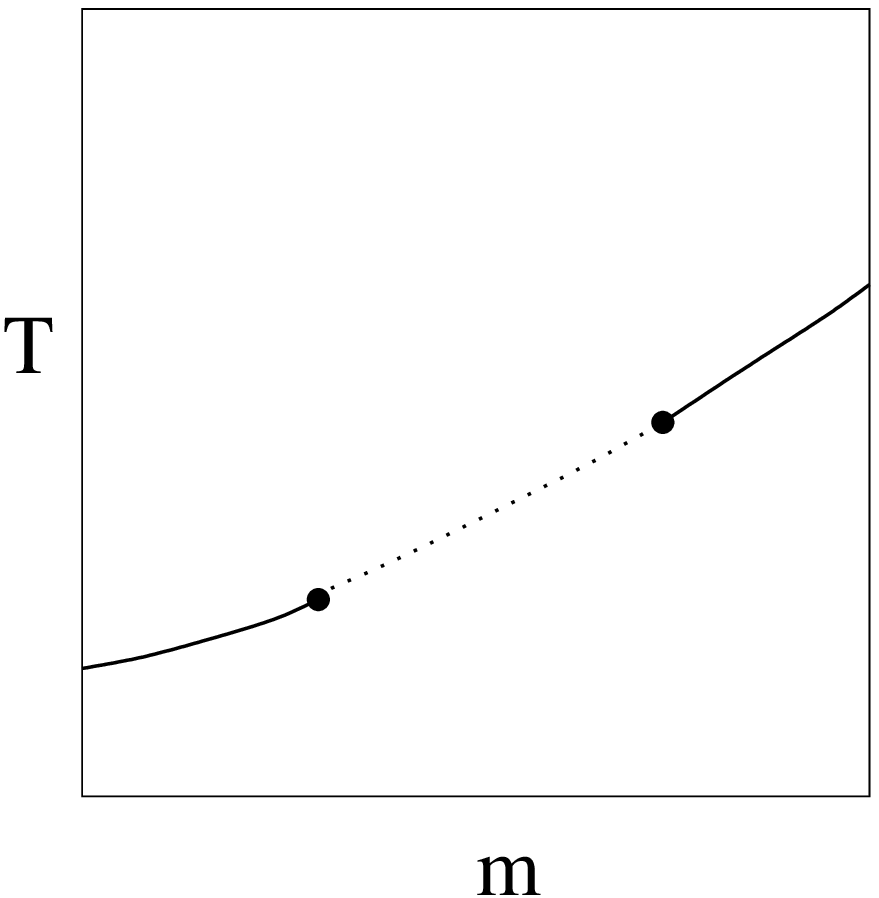}
\put(-85,80){$\langle {\rm Im}(L)\rangle\neq 0$}
\put(-55,25){$\langle {\rm Im}(L)\rangle =0$}
\caption[]{Left: phase diagram for imaginary $\mu$. Vertical lines are
first order transitions between different $Z(3)$-sectors, arrows indicate the phase
of the Polyakov loop. The $\mu=0$ chiral/deconfinement transition continues to imaginary chemical
potential, its order depends on $N_f$ and the quark masses.
Right: phase diagram for $N_f=3$ at fixed $\mu=i\pi T$. Solid lines are lines of triple points
ending in tricritical points, which are connected by a $Z(2)$-line.}
\label{schem}
\end{figure}

Hence, for imaginary $\mu=i\mu_i$ there is a global $Z(3)$ symmetry, even in the 
presence of finite mass quarks. Its spontaneous breaking
implies transitions between neighbouring sectors for the mentioned critical values. 
These are first order phase transitions for high temperatures and analytic crossovers
for low temperatures \cite{rw,fp1,el1}, as shown schematically in \fig\ref{schem} (left).  
For $\mu=i\pi T$ the order parameter to distinguish between the phases is the
imaginary part of the Polyakov loop, ${\rm Im}(L)$. At high temperature there is a two-phase 
coexistence with fluctuations about $\langle{\rm Im}(L)\rangle=\pm \sqrt{3}/2$, 
on either side of the boundary.
At low temperatures ${\rm Im}(L)$ fluctuates smoothly between those values.
The situation is identical for the other critical values of $\mu_i$, up to $Z(3)$ rotations.

Away from a critical value of $\mu_i$, there is a chiral/deconfinement
transition line separating high and low temperature regions. This line represents the analytic
continuation of the chiral/deconfinement  transition at real $\mu$. Its nature
depends on the number of quark flavours and masses.
Early evidence \cite{fp1,el1} is consistent with this line meeting the $Z(3)$ transition at its endpoint 
between first order and crossover, and our present analysis unambiguously confirms this. 
The nature of the endpoint of the $Z(3)$ transition line has already been investigated for $N_f=4$ \cite{el2}
and more recently for $N_f=2$ \cite{mass}.  

In this letter we study the nature of this junction at fixed $\mu=i\pi T$
in $N_f=3$ QCD as a function of quark mass. There are three possibilities, which we
shall find to be all realised. For small masses the chiral transition is first
order and branches off the $Z(3)$-transition line, rendering the meeting point of the three
first order lines a triple point. For intermediate masses, the $Z(3)$ transition ends in a 
second order endpoint with 3d Ising 
universality, i.e.~the chiral/deconfinement transition
in its vicinity is a crossover. For large masses there is a first order deconfinement transition
meeting the $Z(3)$ transition again in a triple point.  
Hence, for fixed $\mu=i\pi T$, we obtain a $(T,m)$ phase diagram as in \fig\ref{schem} (right). The 
endpoints of the solid lines, which separate triple points from Ising points, correspond to tricritical points.

To establish the phase diagram \fig\ref{schem} (right) numerically, 
we work on lattices with temporal extent $N_t=4$ with standard staggered fermions at fixed $\mu/T=i\pi$, 
using the RHMC algorithm and setting aside possible issues with taking 
a fractional power of the fermion determinant.
For fixed $N_t$, temperature is tuned by varying
the lattice gauge coupling $\beta$. Hence, for a given bare quark mass $am$, we 
investigate the nature of the transition as a function of $\beta$.  
To determine this, we analyse the finite size scaling
of the Binder cumulant 
\beq 
B_4(X) \equiv \langle (X - \langle X \rangle)^4 \rangle 
/ \langle (X - \langle X \rangle)^2 \rangle^2,
\eeq
with $X={\rm Im}(L)$ and $\langle X \rangle = 0$.  
For $\mu/T=i\pi$, every $\beta$-value represents a point on the phase boundary and thus
is pseudo-critical. In the thermodynamic limit, $B_4(\beta)=3, 1.5, 1.604, 2$ for crossover, 
first order triple point, 3d Ising and tricritical transitions, respectively. 
On finite $L^3$ volumes the steps between these
values are smeared out to continuous functions whose gradients increase with volume.
The critical coupling $\beta_c$ for the endpoint is obtained as the intersection of curves from
different volumes. In the scaling region around a critical $\beta_c$,  
$B_4$ is a function of $x=(\beta-\beta_c)L^{1/\nu}$ alone and can be expanded as 
\beq
B_4(\beta,L)=B_4(\beta_c,\infty)+ax + b x^2+\ldots,
\label{scale}
\eeq
up to corrections to scaling,
with the critical exponent $\nu$ characterising the approach to the thermodynamic limit.
The relevant values for us are $\nu=1/3, 0.63, 1/2$ when 
$\beta_c$ corresponds to a first order, 3d Ising or tricritical transition, respectively.

\begin{figure}
\vspace*{-0.2cm}
\hspace*{-1.1cm}
\includegraphics[width=0.37\textwidth]{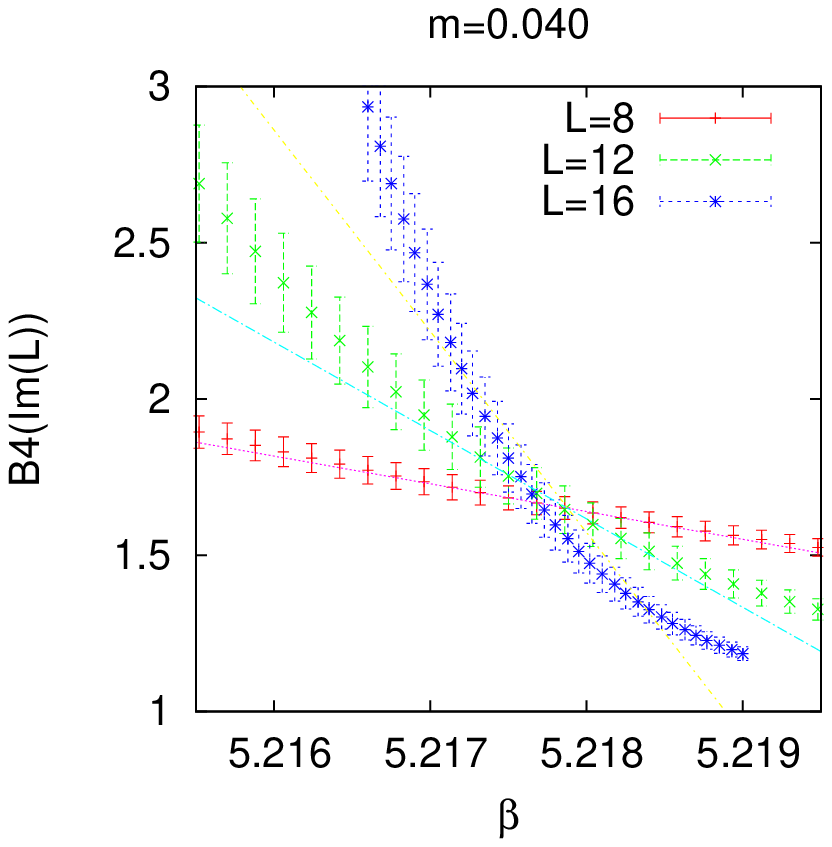}\hspace*{-1.9cm}
\includegraphics[width=0.37\textwidth]{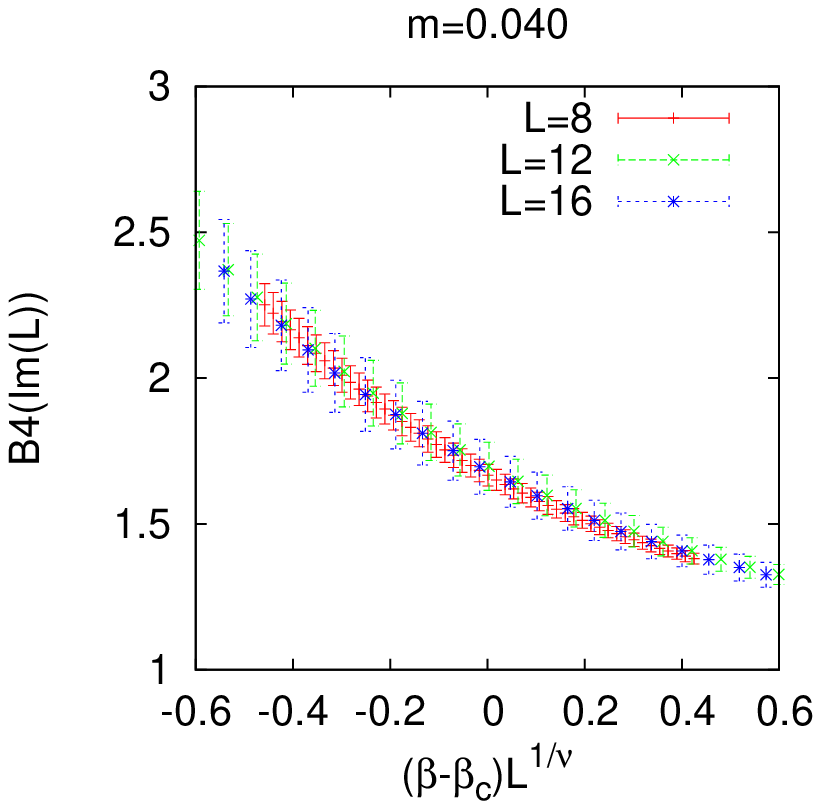}
\vspace*{-0.4cm}
\caption[]{Finite size scaling of $B_4$ for a small quark mass. On the right, the 
critical exponent was fixed to $\nu=1/3$, 
corresponding to a first order transition.}
\label{b4}
\end{figure}

For each quark mass, we simulated lattices of sizes $L=8, 10, 12$ (20 in a few cases), at typically 8-14 different $\beta$-values,
calculated $B_4({\rm Im}(L))$ and 
filled in additional points by Ferrenberg-Swendsen reweighting \cite{fs}.
\fig\ref{b4} shows an example for a light quark mass $am=0.04$.
$B_4$ moves from large values (crossover) at low $\beta$-values (i.e.~low $T$)
towards 1 (first order transition) at large $\beta$-values (i.e.~high $T$). 
In the neighbourhood of the intersection 
point, we then fit all curves simultaneously to
\eq(\ref{scale}), thus extracting $\beta_c,B_4(\beta_c,\infty),\nu, a,b$.
We observe that the value of the Binder cumulant at the intersection can be far from the expected
universal values in the thermodynamic limit. This is a common situation:
large finite-size corrections are observed in simpler spin models even
when the transition is strongly first-order \cite{1st}. Moreover, in our case,
logarithmic scaling corrections will occur near a tricritical point since
$d=3$ is the upper critical dimension in this case \cite{ls}.
Fortunately, the critical exponent $\nu$, which determines the 
approach to the thermodynamic limit, is less sensitive to
finite-size corrections and in \fig\ref{b4} consistent with $\nu=1/3$, its value at a first order
transition. A check is to fix $\nu$ to one of the universal values 
and see whether the curves collapse under the appropriate rescaling, \fig\ref{b4} (right). 
Note that the critical coupling determined from the intersection of the
$B_4$ curves \fig\ref{b4} (left) is consistent with that extracted from the
peak of the specific heat or the chiral susceptibility.

\begin{figure}[t]
\hspace*{-0.9cm}
\includegraphics[width=0.31\textwidth]{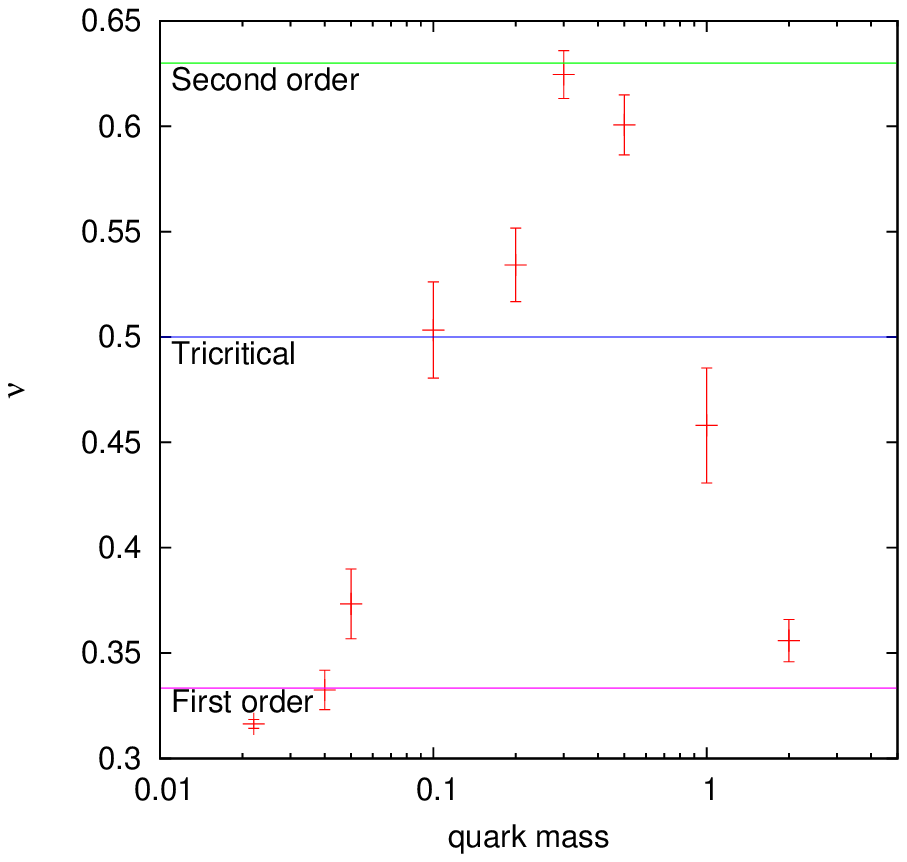}\hspace*{-1.2cm}
\includegraphics[width=0.31\textwidth]{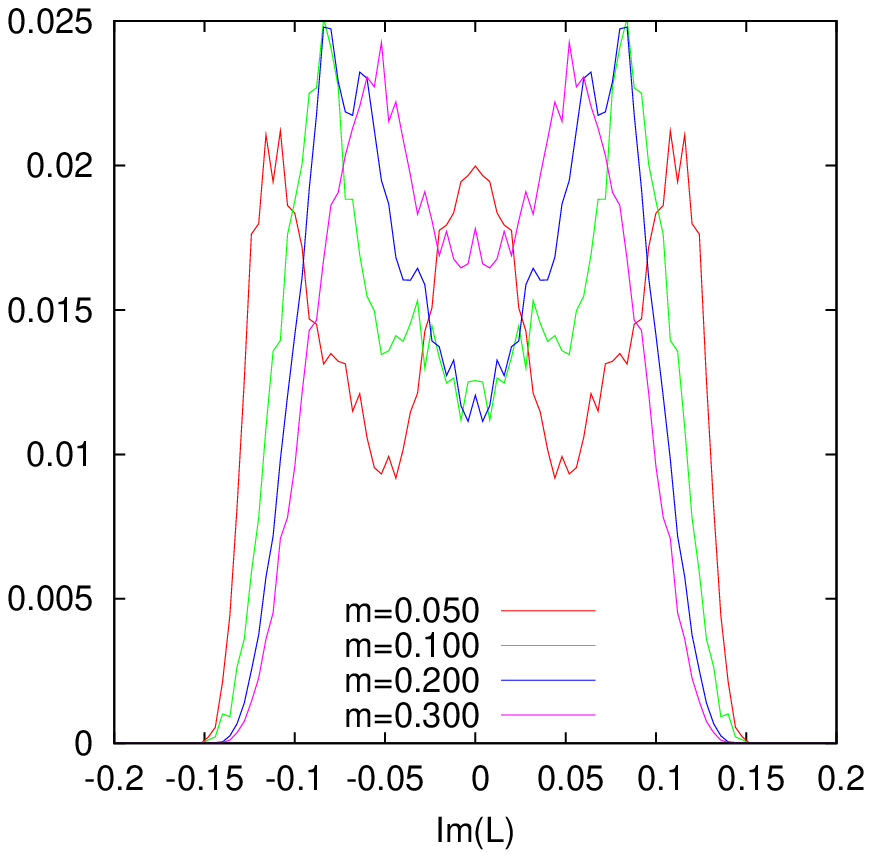}
\caption[]{Left: Critical exponent $\nu$ at $\mu/T=i\pi$. Right: Distribution of ${\rm Im}(L)$ at the
endpoint of the $Z(3)$ transition.}
\label{exp}
\end{figure}

Proceeding in this way, we have investigated quark masses
distributed over a large range, with results summarised in \fig\ref{exp} (left).
We find unambiguous evidence for a change from first order scaling to 3d Ising scaling, 
and back to first order scaling as the quark mass is made larger. 
Note that, in the infinite volume limit, the curve would be replaced by a non-analytic step function, whereas the smoothed-out rise and fall in \fig\ref{exp} (left) corresponds to finite volume corrections.

The results from the finite size scaling of $B_4$ can be further sharpened by looking at 
the probability distribution of ${\rm Im}(L)$ at the critical couplings $\beta_c$, 
corresponding to the crossing points. This is shown in \fig\ref{exp} (right) for 
masses $am=0.05,0.1,0.2,0.3$ for $L=16$.
The lightest mass  
displays a clear three-peak structure, indicating
coexistence of three states 
at the coupling 
$\beta_c$, which therefore corresponds to a triple point. The same observation holds for 
heavy masses. For $am=0.1,0.2$ the central peak is disappearing and 
for $am=0.3$ we are left with the two peaks characteristic for the magnetic direction 
of 3d Ising universality.  

Hence, for small and large masses, we have unambiguous evidence that
the boundary point between a first order $Z(3)$ transition and a crossover at $\mu=i\pi T$ corresponds to a triple point. 
This implies that two additional first order lines branch off the $Z(3)$-transition line as in 
\fig\ref{schem} (left), which are to be identified as the chiral (for light quarks) or deconfinement (for heavy 
quarks) transition at imaginary chemical potential. 
This is expected on theoretical grounds: for $m=0$ or $+\infty$, these 
transitions are first-order for any chemical potential.
The fact that the endpoint of the $Z(3)$ transition line changes its nature from a triple point at 
low and high masses to 
second order for intermediate masses implies the existence of two tricritical points. 
Our current data on $N_t=4$ put these between
$0.07<am_{\rm tric 1}<0.3$ and $0.5< am_{\rm tric 2}<1.5 $.

Since our $N_t=4$ lattice is very coarse, $a\sim 0.3$ fm, an important question concerns 
cut-off effects. These strongly affect quark masses, and thus in particular
the values of the tricritical quark masses where the changes from a triple point to
a critical Ising point happen. 
However, universality implies that critical
behaviour is insensitive to the cut-off, as long as the global symmetries 
of the theory are not changed. 
Our calculation is therefore 
sufficient to establish the qualitative picture \fig\ref{schem} (right) for the continuum phase diagram at 
$\mu=i(2n+1)\pi T/3 $ for three
degenerate flavours.

\begin{figure}[t!!!]
\includegraphics[width=0.25\textwidth]{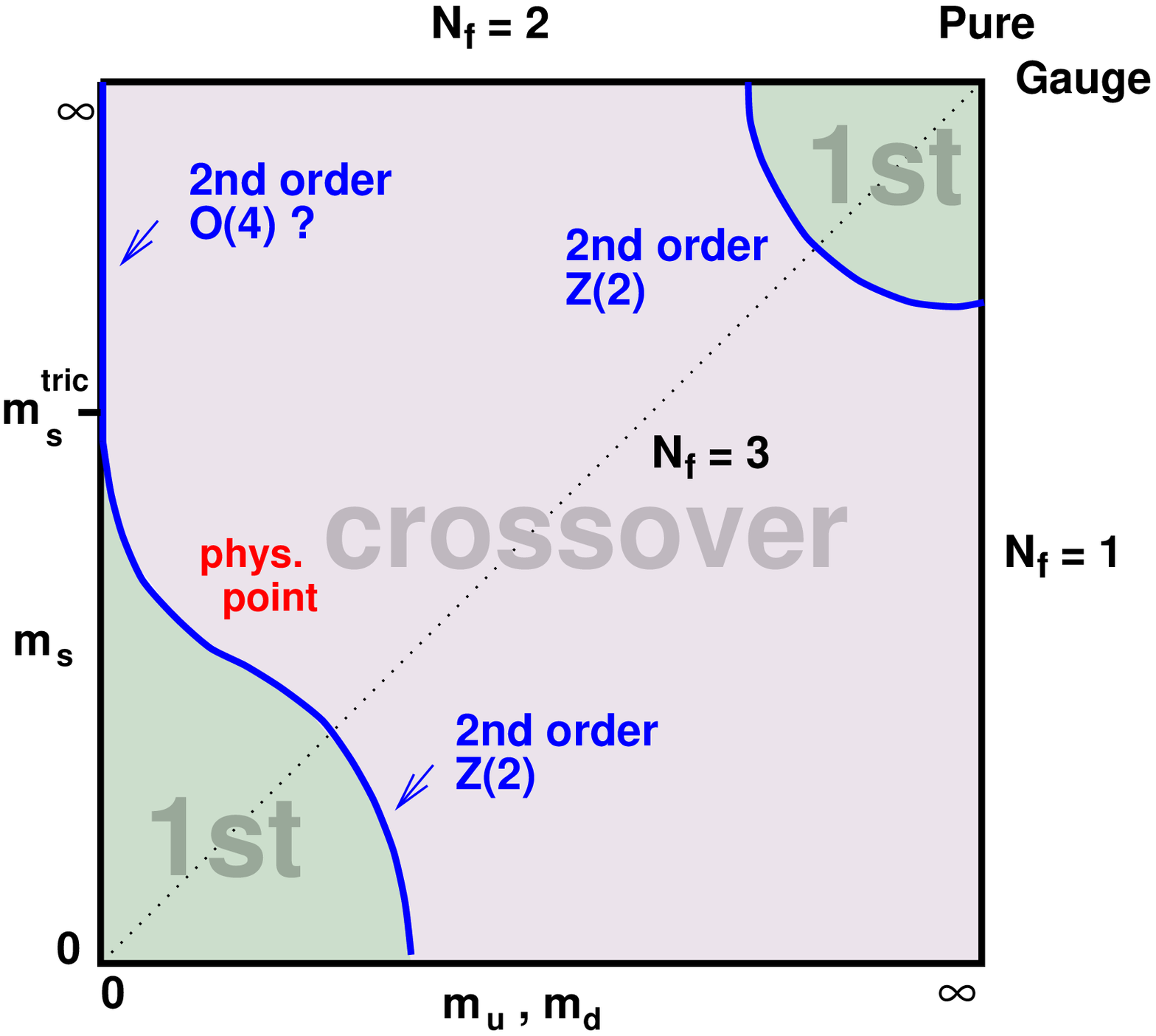}
\includegraphics[width=0.22\textwidth]{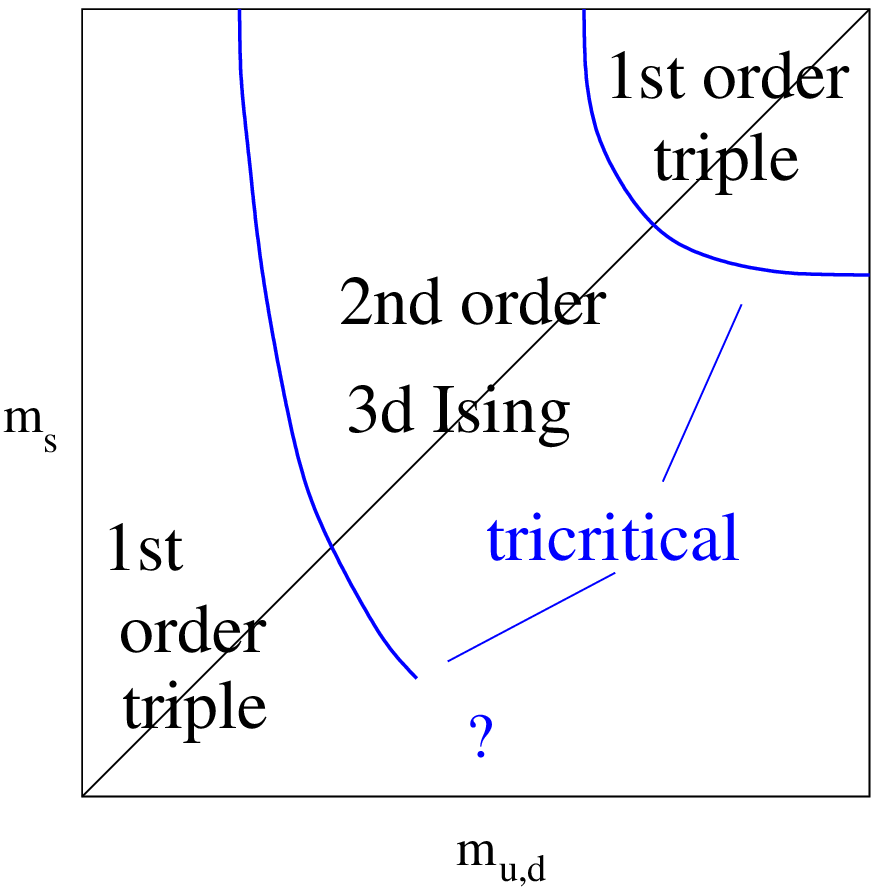}
\caption[]{Order of the transition as a function of quark masses.
Left: quark hadron transition at $\mu=0$.
Right: the $Z(3)$-transition endpoint at $\mu/T=i\pi/3$.}
\label{col}
\end{figure}
Let us now discuss how this critical structure is embedded in the 
parameter space with non-degenerate quark masses, \fig\ref{col} (right).
The case $N_f=3$ corresponds to the diagonal, 
with two tricritical points separating triple points from second order points.
With non-degenerate quark masses these tricritical points will trace out tricritical lines,
$m^{\rm tric}_s(m_{u,d})$.
In the case of heavy quarks the situation is qualitatively 
the same for any $N_f=1,2,3$ \cite{Nf=1,lp}.
In the light quark regime, there is an interplay between the $Z(3)$ and
chiral symmetries and the situation may be more complicated. The findings reported in 
\cite{mass} imply the existence of a finite tricritical light quark mass also for $N_f=2$.
It would then seem
natural that the tricritical points for $N_f=2,3$ are continuously connected by varying the strange quark 
mass, though this needs to be confirmed by explicit calculations.
We stress that all critical structure
indicated in \fig\ref{col} (right) 
can be determined reliably with standard Monte Carlo techniques, and continuum
extrapolations are feasible with current resources. Knowledge of this phase diagram in the continuum
should provide valuable benchmarks for the description of the QCD phase diagram by effective models.

In order to establish the connection between imaginary and real chemical potential, let us briefly recall the situation at $\mu=0$, \fig\ref{col} (left).
The deconfinement transition in pure gauge theory 
is first order. In the presence of dynamical quarks, 
it weakens with decreasing quark mass until it disappears along the deconfinement 
critical line with 3d Ising universality. The critical point for $N_f=1$ was 
determined in \cite{Nf=1}, and more recently for $N_f=1, 2, 3$ in a strong
coupling expansion for a coarse $N_t=1$ lattice~\cite{lp}.
Similarly, the chiral transition for $N_f=2+1$ is first order and weakens with increasing quark mass,
until it disappears at a chiral critical line with 3d Ising universality \cite{kls,fp3}.
When a chemical potential is switched on, these critical lines sweep out critical surfaces which
continue in the imaginary $\mu$ (or $-\mu^2$) direction and join the tricritical lines
at $\mu=i\pi T/3$. 
We shall now illustrate this for the deconfinement critical surface.

\begin{figure}[t!!!]
\hspace*{-0.9cm}
\includegraphics[width=0.26\textwidth]{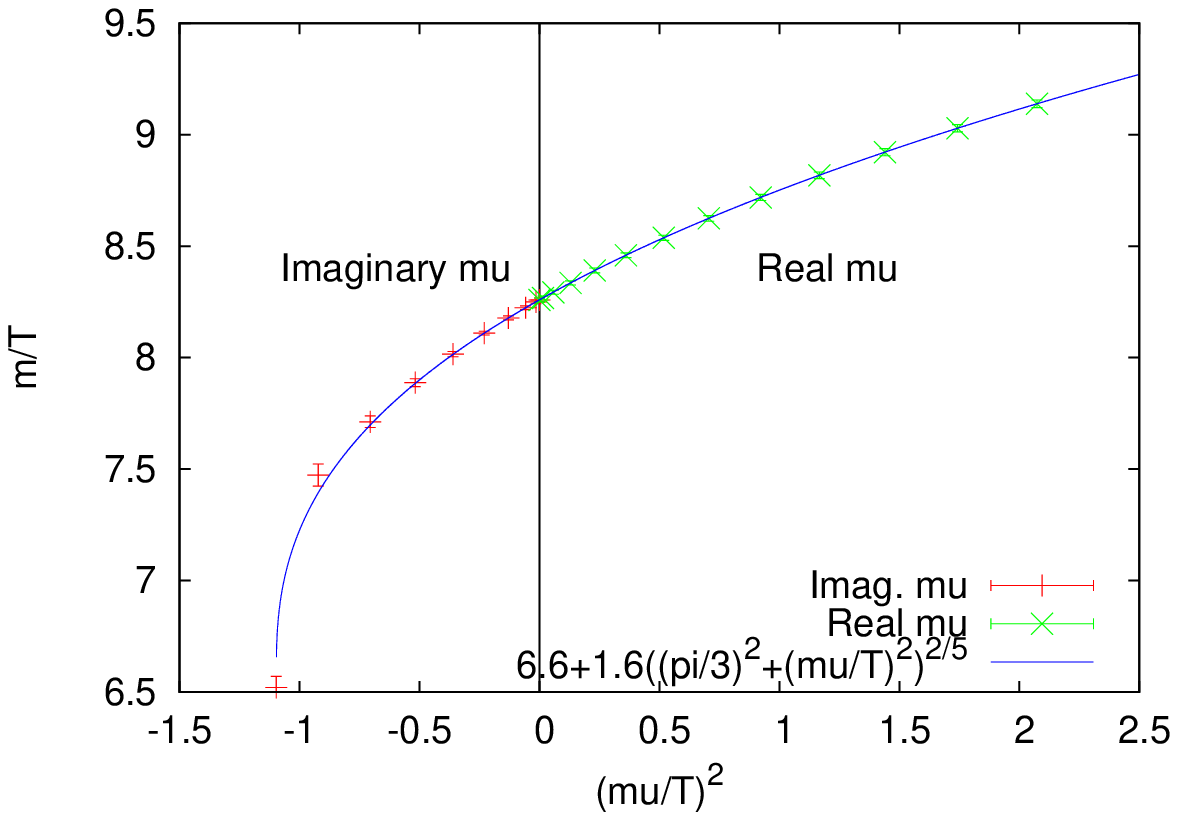}
\includegraphics[width=0.26\textwidth]{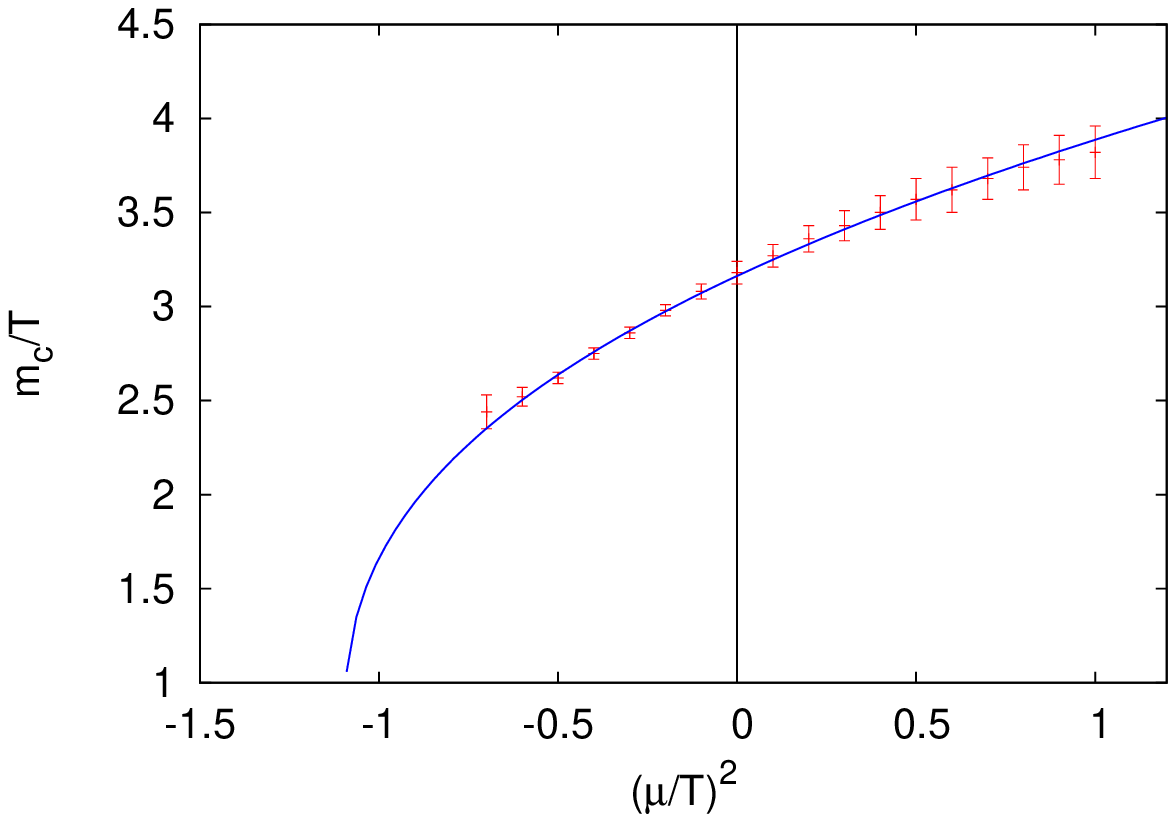}
\vspace*{-0.4cm}
\caption[]{Critical line $m_c(\mu^2)$ in the 3-state Potts model~\cite{fkt} (left) and
for QCD in a strong 
coupling expansion~\cite{lp} (right).}
\label{tric}
\end{figure}

By universality, the properties of a second order transition are the same in a model sharing the same global symmetry, and for the deconfinement transition this is the 3d three-state Potts 
model with Hamiltonian
\beq
H=-k\sum_{i,\bx}\delta_{\phi(\bx),\phi(\bx+\hat{i})}-\sum_\bx [h \phi(\bx)+h'\phi^*(\bx)]\;,
\eeq 
where $\phi(\bx)$ is a $Z(3)$-spin, 
and the couplings are identified as
$h=\exp-(M-\mu)/T, h'=\exp-(M+\mu)/T$, while $k$ increases with temperature. 
The qualitative change of the critical deconfinement
line with chemical potential can be calculated in this model and one finds the first order region
to shrink with real $\mu$ \cite{fkt}. 
The same observation is made for full QCD on a coarse lattice with a strong coupling expansion \cite{lp}.
See \fig\ref{tric}. 

Both calculations show the continuation of the critical mass to negative $\mu^2$ and 
a non-analyticity when joining the $Z(3)$-transition at $\mu=i\pi T/3$. 
However, in both cases it has not been fully realised that this junction is tricritical.
Since chemical potential enters the partition 
function of the Potts model in the same way as in  QCD, it features the same $Z(3)$-transitions.  We can therefore directly check our QCD 
results in the heavy mass
region against those in the Potts model at the same value $\mu=i\pi T/3$. For the latter, the Binder 
cumulant of the spin magnetisation was measured \cite{fkt}. 
We reanalysed those data fitting them to 
the scaling form, \eq(\ref{scale}), and indeed find a change from first order behaviour, 
$\nu=0.33$, at large values of $M/T$  to second order 3d Ising, $\nu=0.63$, implying again
a tricritical point. 

Quite generally, a tricritical point
represents the confluence of two ordinary critical points, i.e.~in the heavy mass region 
the critical endpoints of the deconfinement transition at $\mu=i\pi T/3 (1\pm\varepsilon)$ merging into the $Z(3)$ endpoint at $\mu=i\pi T/3$.
The deviation from the symmetry plane, $((\mu/T)^2+(\pi/3)^2)$,
corresponds to an external field in a spin model, and the way a critical line leaves a tricritical point
in an external field is again governed by universality \cite{ls},
\beq
\frac{m_c}{T}(\mu^2)=\frac{m_c}{T}(0) + K \left[\left(\frac{\pi}{3}\right)^2+\left(\frac{\mu}{T}\right)^2\right]^{2/5}\;.
\eeq
\fig\ref{tric} shows that the data from \cite{fkt} and \cite{lp} can both be fitted to this form. 
An excellent description of the data is achieved, reaching
far into the real chemical potential region. 
Thus we conclude that for heavy quark masses, the form of the critical
surface of the deconfinement transition is  determined by tricritical scaling of the $Z(3)$ transition at imaginary $\mu=i\pi T/3$.

It is clear that the chiral critical surface will likewise 
terminate on the chiral tricritical line at $\mu=i\pi T/3$.
Unfortunately, for the chiral critical surface no suitable effective model for finite 
density is available and we presently do not know
whether tricritical scaling also shapes the chiral critical surface. 
This could be answered conclusively by extensive simulations at intermediate values of $\mu_i$.
However, we do find $m_c(\mu=i\pi T/3)>m_c(0)$, which indicates a monotonous reduction
of the critical quark mass as $\mu^2$ is increased. This independently confirms
the weakening of the first order chiral transition with
real chemical potential observed previously in \cite{fp1,fp3,fp4}.



\begin{thebibliography}{99}


\bibitem{review}
O.~Philipsen,
  PoS {\bf LAT2005} (2006) 016
  [PoS {\bf JHW2005} (2006) 012]
  [arXiv:hep-lat/0510077].
\bibitem{fp1}
P.~de Forcrand and O.~Philipsen,
  Nucl.\ Phys.\  B {\bf 642}, 290 (2002)
  [arXiv:hep-lat/0205016].
\bibitem{el1}
M.~D'Elia and M.~P.~Lombardo,
  Phys.\ Rev.\  D {\bf 67}, 014505 (2003)
  [arXiv:hep-lat/0209146].
\bibitem{rw}
A.~Roberge and N.~Weiss,
  Nucl.\ Phys.\  B {\bf 275}, 734 (1986).
\bibitem{mass}
M.~D'Elia and F.~Sanfilippo,
  arXiv:0909.0254 [hep-lat].
\bibitem{el2}
M.~D'Elia, F.~Di Renzo and M.~P.~Lombardo,
  Phys.\ Rev.\  D {\bf 76}, 114509 (2007)
  [arXiv:0705.3814 [hep-lat]].
\bibitem{fs}
A.~M.~Ferrenberg and R.~H.~Swendsen,
  Phys.\ Rev.\ Lett.\  {\bf 63}, 1195 (1989).
\bibitem{1st}
A.~Billoire, T.~Neuhaus and B.~Berg,
  Nucl.\ Phys.\  B {\bf 396}, 779 (1993)
  [arXiv:hep-lat/9211014].
\bibitem{ls}
I.D.~Lawrie and S.~Sarbach, in {\it Phase transitions and critical phenomena}, 
eds. C.~Domb and J.L.Lebowitz, vol.9, 1 (1984).  
\bibitem{lp}
J.~Langelage and O.~Philipsen,
  arXiv:0911.2577 [hep-lat].
\bibitem{Nf=1}
  C.~Alexandrou et al., 
  Phys.\ Rev.\  D {\bf 60} (1999) 034504
  [arXiv:hep-lat/9811028].
\bibitem{kls}
F.~Karsch, E.~Laermann and C.~Schmidt,
  Phys.\ Lett.\  B {\bf 520}, 41 (2001)
  [arXiv:hep-lat/0107020].
\bibitem{fp3}
P.~de Forcrand and O.~Philipsen,
  JHEP {\bf 0701}, 077 (2007)
  [arXiv:hep-lat/0607017].
\bibitem{fkt}
S.~Kim, Ph.~de Forcrand, S.~Kratochvila and T.~Takaishi,
  PoS {\bf LAT2005}, 166 (2006)
  [arXiv:hep-lat/0510069].
\bibitem{fp4}
  P.~de Forcrand and O.~Philipsen,
  JHEP {\bf 0811} (2008) 012
  [arXiv:0808.1096 [hep-lat]].
  
\end{thebibliography}
\end{document}